\documentclass {aa}
\usepackage{psfig}

\def\cgs{{erg cm$^{-2}$ s$^{-1}$}}
\def\xnu{{$\chi^{2}(\nu)$}}
\begin{document}

\title{Exploring the spectral properties of faint \\hard X-ray sources with {\it XMM$-$Newton}}

\author{E.~Piconcelli\inst{1,2}, M.~Cappi\inst{1}, L.~Bassani\inst{1}, F.~Fiore\inst{3}, G.~Di~Cocco\inst{1}, J.~B.~Stephen\inst{1}}
\institute{IASF/CNR, via Piero Gobetti 101, I--40129 Bologna,
             Italy \and Dipartimento di Astronomia, Universit\`a di
             Bologna, via Ranzani 1, I--40127 Bologna, Italy \and
              Osservatorio Astronomico di Roma, Via Frascati 33, I--00044 Monteporzio, Italy}
\authorrunning{E. ~Piconcelli  et al.}
\titlerunning{ An {\it XMM$-$Newton} Spectral Survey of Faint Hard X-ray Sources}
\offprints{Enrico~Piconcelli, e-mail: piconcel@tesre.bo.cnr.it}
\date{Received / accepted}

\abstract{
We present a spectroscopic study of 41 hard X-ray sources detected serendipitously
with high significance ($>$5$\sigma$ in the 2-10 keV band) in seven {\it EPIC} 
performance/verification phase observations. 
The large collecting area of {\it EPIC} allows us to explore the spectral properties of these
 faint hard X-ray sources with 2$<$$F_{2-10}$$<$80 $\times$ 10$^{-14}$ \cgs~ 
even though the length of the exposures are modest ($\sim$20 ks). 
Optical identifications are available for 21 sources of our sample.
Using a simple power law plus Galactic absorption model
we find an average value of the photon index $\Gamma$$\sim$1.6-1.7, broadly consistent with
 recent measurements made at similar fluxes with {\it ASCA} and with {\it Chandra} stacked spectral analyses.
We find that 31 out of 41 sources are well fitted by this simple model and only eight sources require absorption 
in excess of the Galactic value. Interestingly enough, one third of these absorbed sources are broad line objects, though with moderate column densities.
Two sources in the sample are X-ray bright optically quiet galaxies and show flat X-ray spectra. 
Comparing our observational results with those expected from standard synthesis models of the cosmic X-ray background (CXB)
we find a fraction of unabsorbed to absorbed sources larger than predicted by theoretical models
at our completeness limit of $F_{2-10}$$\sim$5 $\times$ 10$^{-14}$ \cgs.
The results presented here illustrate well how wide-angle surveys performed 
with {\it EPIC} on board {\it XMM-Newton} allow population studies of interesting and unusual sources  to be made
as well as enabling constraints to be placed on some input parameters for synthesis models 
of the CXB.

 }
\maketitle
%
%

\section{Introduction}
It is now widely accepted that the bulk of the Cosmic X-ray background (CXB)
in the range 0.1-10 keV is the result of the integrated emission of unresolved
sources over cosmic time (Giacconi et al. 2001).
Recent extremely deep surveys by  {\it Chandra} (Tozzi et al. 2001; Brandt et al. 2001a) and {\it XMM-Newton}
 (Hasinger et al. 2001), have resolved up to $\sim$75-90\% of the CXB into discrete sources, 
down to a limiting 2-10 keV flux of $\sim$10$^{-16}$ erg cm$^{-2}$ s$^{-1}$. On-going follow-up optical identification
programs suggest that most of these sources are AGNs, while a sizeable fraction of the
rest are optically faint ($I$$>$24) objects ($\sim$30\%), likely to be highly obscured AGNs
 (Alexander et al. 2001) or luminous bulge-dominated galaxies (Barger et al. 2001, Cowie et al. 2001).
Large area surveys (e.g. from {\it ASCA} and {\it BeppoSAX}; Akiyama et al. 2000;
Della Ceca et al. 1999; Giommi, Perri \& Fiore 2000), limited to higher fluxes
but providing a larger number of sources, confirm the above results.
At a flux limit of $\sim$5 $\times$ 10$^{-14}$ erg cm$^{-2}$ s$^{-1}$ in the range 5-10 keV, the bulk of 
the sources are broad line AGNs, 
while most of the others are classified as
Seyfert 1.8-2 galaxies and optically 'red' quasars (La Franca et al. 2000).
Surprisingly, {\it Chandra} serendipitously discovered that also some
normal galaxies, without evidence for any activity (i.e. AGN and/or starburst) in the optical band, are
X-ray loud (Fiore et al. 2000).

To date very few X-ray spectra of such faint 
sources are reported in the literature (but see pioneering results from Vignali et al. 2000; Sakano et al. 1998
and Crawford et al. 2001). Such spectral information is crucial in order 
to accurately determine source-by-source the amount of absorption, the spectral shape
and the presence of reprocessed features (see e.g. Yaqoob 2000).
 These can also be used as input parameters for the synthesis models of the CXB in
order to evaluate the contribution of the various AGN types and constrain their evolution (Comastri et al. 2001; 
Gilli, Salvati \& Hasinger 2001). 
Important questions are still unanswered, such as the relationship between optical identification and 
X-ray characteristics as well as the role of absorption in the different types of active galaxies
 (Salvati \& Maiolino 2000; Pappa et al. 2001a).
There is growing evidence that not only narrow line but also  broad line AGNs suffer from intrinsic absorption 
(Risaliti et al. 2000), which is present both in radio-quiet  (Gallagher et al. 2000; 
Brandt al. 2001b; Maloney \& Reynolds 2001; Collinge \& Brandt 2000) and in radio-loud 
objects (Cappi et al. 1997; Yuan et al. 2000; Sambruna, Eracleous \& Mushotzky 1999) 
at all redshifts. These obscured sources show hard spectral indices which match well with the CXB slope in 
the hard X-ray band ($\sim$1.4, e.g. Marshall et al. 1980).
Moreover doubts are emerging about the existence of the long sought after high luminosity type 2 AGNs (Halpern,Turner \& George 1999),
 the so-called QSO 2s, which should play
a significant (and perhaps a major) role in the production of the hard CXB (Fabian \& Iwasawa 1999): 
from recent deep observations only two cases have  
been reported (Norman et al. 2001 and Stern et al. 2001). It would appear to be difficult to discover a
large number of distant Compton-thick QSOs with present X-ray observatories (Fabian, Wilman \& Crawford 2001).

Based on the above arguments, we have started an extensive program with {\it XMM-Newton}
aimed at studying the brightest (in the hard X-ray band)
sources serendipitously detected in a number of fields with moderate exposure ($\sim$20 ks).
This work has been designed to complement at an intermediate flux level ($\sim$10$^{-14}$ \cgs)
the X-ray population studies made by very deep pencil-beam observations.
On account of its good positional accuracy ($\sim$6 arcsec error radius)
and its unprecedented sensitivity in the 2-10 keV band, the {\it XMM-Newton}
satellite is currently the most appropriate telescope with which to pursue such a study.
\begin{table*}
\caption{Journal of the {\it XMM-Newton} observations.}
\label{tab1}
\begin{center}
\begin{tabular}{lcccccccc}
\hline
\multicolumn{1}{c} {Field} &
\multicolumn{1}{c} {R.A.} &
\multicolumn{1}{c} {Dec.}&
\multicolumn{1}{c} {Orbit}&
\multicolumn{1}{c} {Date} &
\multicolumn{3}{c} {Exposure (s)} &
\multicolumn{1}{c} {Filter} \\
 & & & & &PN &M1 &M2&PN M1 M2\\
\hline\hline\\
PKS0312$-$770  &03 11 55.0 &$-$76 51 52&057 &2000-03-31 &26000 &25000&24000&Tc Tc Tc \\
MS1229.2$+$6430&12 31 32.0 &$+$64 14 21 &082 &2000-05-20 &22900&18600&22900 &Th Th Th\\
IRAS13349$+$2438&13 37 19.0 &$+$24 23 03&097&2000-06-20&$-$ &41300& 38600 &$-$ Me Th\\
Abell 2690&00 00 30.0 &$-$25 07 30&088 &2000-06-01 &21000 &16600 &15300 & Me Me Me\\
MS 0737.9$+$744&07 44 04.5 &$+$74 33 49&063 & 2000-04-12& 15000 &17800 &26100& Th Th Th\\
Markarian 205& 12 21 44.0 &$+$75 18 37&075&2000-05-07&17000 &$-$ &14800& Me $-$ Me\\
Abell 1835&14 01 02.0 &$+$02 52 41 &101 &2000-06-27&22900 &23700 &26400 &Th Th Th\\
\hline 
\end{tabular} 
\end{center}
Optical blocking filters used during observations: Th=thin, Me=medium and Tc=thick.
\end{table*}
\section{ {\it XMM-Newton} observations}
\subsection{Data reduction}
The {\it XMM-Newton} observatory was launched on December 10th 1999
and was placed in a 48 hour orbit (Jansen et al. 2001).
{\it EPIC} ({\it European Photon Imaging Cameras}) on-board {\it XMM-Newton} consists of two 
{\it MOS} CCD front-illuminated arrays
(Turner et al. 2001) and one {\it PN} CCD back-illuminated array (Struder et al. 2001) 
for X-ray imaging and spectroscopy in the range $\sim$0.1-10 keV. 
Each camera is located in the focal plane of one of three X-ray grazing
incidence telescopes and provides a 30 arcmin diameter field of view.
{\it EPIC} is equipped with a filter wheel which  carries three different
blocking filters, i.e. Thin (40 nm Al), Medium (80 nm Al) and 
Thick (200 nm Al), optimised for different investigations.

 In the present study we investigate the {\it EPIC} fields of seven observations performed 
during the performance/verification phase. These fields are listed in 
Table 1 together with their observational details. They cover in total an area of $\sim$1.4 deg$^{2}$ 
of the sky and they all have {\it b}$\geq$30 deg so that we can exclude heavy contamination 
from Galactic sources and high Galactic column densities along the line of sight. 
We are able to reach 4-5 arcsec absolute pointing accuracy 
reconstruction for the boresight instruments using the Attitude History 
Files (AHFs) available for all datasets, as demonstrated by the 
{\it EPIC} calibrations (for more details see
 http://xmm.vilspa.esa.es/users/calib$_{-}$top.html).

The raw {\it EPIC} observation data files (ODFs) are reduced and 
analysed using the standard Science Analysis System ($SAS$) software package
 (version 5.0, released in 2000, December 15th;  Gondoin 2001). 
We use the {\small EPCHAIN} and {\small EMCHAIN} tasks for the pipeline 
processing of the ODFs to generate the corresponding event files.
These tasks also allow dead and hot pixels to be removed.
On each event file we then apply an exposure time correction in order to exclude 
intervals contaminated by soft proton flares:  
we extract  background lightcurves at energy $>$10 keV to identify 
and remove periods of high count rates (typically 0.2 cts$/$s). 
The resulting clean integration times for 
each {\it EPIC} camera are listed in Table 1. 
We select pixel patterns $\leq$12 and =0 for {\it MOS} and {\it PN} events,
respectively, in order to filter out cosmic-ray tracks and/or spurious noise events 
not created by X-rays.
 We use the {\small XMMSELECT} task to filter data and to  generate 
lightcurves, images and spectra. All our images are binned to obtain 
a pixelsize of 4.4$\times$4.4 arcsec.
\subsection{Sample creation}
We have created a 2-10 keV {\it PN} image of all but one (IRAS13349+2438) of the fields listed in table 1 
which have then been used to detect hard X-ray serendipitous sources. 
The {\it PN} observation of IRAS13349+2438 was carried out in small window mode (Studer et al. 2001) and  
so we have used a 2-10 keV {\it MOS1} image of this field.
We investigate in the 2-10 keV range because  to date 
it is the nearest X-ray spectroscopically accessible band to the CXB $\nu$$F_{\nu}$ peak at $\sim$30 keV.
The main reason for which we choose the {\it PN} camera to perform such a survey is its greater effective area
with respect to the {\it MOS} (in particular in the hard band it is a factor $\sim$2.5 times at 8 keV), which enables us
 to be more accurate in the detection of faint hard X-ray sources.

In the following we
briefly summarize our detection procedure.
For each 2-10 keV image we generate an exposure map containing 
information about filter transmission and spatial quantum efficiency
 which has been used by other $SAS$ tasks of the detection chain.
The area of the image where source searching is performed
is marked by a detection mask created by the {\small EMASK} task.
We run the {\small EBOXDETECT} task, the standard $SAS$ sliding box 
cell detection algorithm, with local background subtraction (i.e. 'local mode')  
in order to have a list of candidate sources. This list is given as input to 
 the {\small ESPLINEMAP} task to produce a smooth background map. 
This background map is then used in a second {\small EBOXDETECT} procedure (i.e.
'map mode').  
We choose $L$=--ln($P$)=15 as the minimum detection likelihood value corresponding 
 to a probability of Poissonian random fluctuations of the counts in the detection cell  
of $P$=3.0 $\times$ 10$^{-7}$ (roughly $\sim$5$\sigma$).
{\small EBOXDETECT} generates a list of selected sources with their
centroid coordinates $(x,y)$. Finally we include in our sample only those sources detected
with at least 35 net counts in a 5$\times$5 pixel box centered on $(x,y)$.
These counts are raw and they must be corrected for both the energy encircled fraction (EEF) and the vignetting factor: in this way
 our selection criterion (i.e. 35 counts) corresponds, for example, to a corrected final flux of $F_{2-10}\sim$2.3 $\times$ 
10$^{-14}$ \cgs~ 
for an on-axis source and to  $F_{2-10}\sim$5 $\times$ 10$^{-14}$ \cgs~ for  a 10 arcmin off-axis source in our shortest {\it PN} exposure 
(i.e. 15 ks, see Table 1).
As the {\it PN} and {\it MOS} FOVs differ slightly we also perform the detection procedure
 described above on the {\it MOS} images in order to detect those sources which lie in the gaps 
between the {\it PN} chips or out of the {\it PN} FOV.
The selection criterion remains the same, except for the fact that in this case we accept only sources having at
 least 20 net counts in a 5$\times$5 pixel box centered on $(x,y)$. In this way we include a further 4
 sources (2 in the gaps and 2 out of the FOV) which were previously undetected.
In total we obtain a sample of 41 hard X-ray selected serendipitous sources. They are listed in Table 2. 
\begin{table*}
\caption{List of the 41 X-ray serendipitous sources included in our sample.}
\label{tab2}
\begin{center}
\begin{tabular}{cccccc}
\hline
\multicolumn{1}{c} {N} &
\multicolumn{1}{c} {Source name} &
\multicolumn{1}{c} {R.A.}&
\multicolumn{1}{c} {Declination}&
\multicolumn{1}{c} {z} &
\multicolumn{1}{c} {Classification}\\ 
  & &(J2000) & (J2000)   && \\
\hline
\hline
\multicolumn{6}{c}{PKS 0312-770 field}\\
1&CXOUJ031015.9-765131&03 10 15.3& $-$76 51 32 &1.187&BL AGN$^{a}$\\
2&CXOUJ031209.2-765213&03 12 08.7&$-$76 52 11 &0.89&BL AGN$^{a}$\\
3&CXOUJ031238.9-765134&03 12 38.8&$-$76 51 31&0.159&Galaxy$^{a}$\\
4&CXOUJ031253.8-765415&03 12 53.5 &$-$76 54 13 &0.683&Red QSO$^{a}$\\
5&CXOUJ031312.1-765431&03 13 11.5 &$-$76 54 28 &1.124&BL AGN$^{a}$\\
6&CXOUJ031314.5-765557&03 13 14.2 &$-$76 55 54 &0.42 &BL AGN$^{a}$\\
7&XMMUJ030911.9-765824&03 09 11.6 &$-$76 58 24&0.268 &Sey 2$^{b}$\\
8&XMMUJ031049.6-763901&03 10 49.5 &$-$76 39 01& 0.380& BL AGN$^{b}$\\
9&XMMUJ031105.1-765156&03 11 05.1 &$-$76 51 56 &$-$ &No cl.\\
\multicolumn{6}{c}{MS1229.2+6430 field}\\
10&XMMUJ123110.6+641851&12 31 10.6 &$+$64 18 51 &$-$ &No cl.\\
11&XMMUJ123116.3+641114&12 31 16.3 &$+$64 11 14 &$-$ &No cl.\\
12&XMMUJ123218.6+640309&12 32 18.6 &$+$64 03 09 &$-$ &No cl. \\
13&XMMUJ123214.2+640459&12 32 14.2 &$+$64 04 59 &$-$ &No cl. \\
14&XMMUJ123013.4+642505&12 30 13.4 &$+$64 25 05 &$-$ &No cl. \\
15&XMMUJ123049.9+640845&12 30 49.9 &$+$64 08 45 &$-$ &No cl. \\
16&XMMUJ123058.5+641726&12 30 58.5 &$+$64 17 26 &$-$ &No cl. \\
\multicolumn{6}{c}{IRAS13349+2438 field}\\
17&XMMUJ133730.8+242305 &13 37 30.8 &$+$24 23 05 &$-$ &No cl. \\
18&XMMUJ133649.3+242004 &13 36 49.3 &$+$24 20 04 &$-$ &No cl. \\
19&XMMUJ133807.4+242411 &13 38 07.4 &$+$24 24 11 &$-$ &No cl. \\
20&XMMUJ133747.4+242728 &13 37 47.4 &$+$24 27 28 &$-$ &No cl. \\
21&XMMUJ133712.6+243252 &13 37 12.6 &$+$24 32 52 &$-$ &No cl. \\
\multicolumn{6}{c}{Abell 2690 field}\\
22&XMMUJ000031.7-255459&00 00 31.7 &$-$25 54 59 &0.283  &BL AGN$^{b}$\\
23&XMMUJ000122.8-250019&00 01 22.8 &$-$25 00 19 &0.968  &BL AGN$^{b}$\\
24&XMMUJ000027.7-250441&00 00 27.7 &$-$25 04 41 &0.335  &BL AGN$^{b}$\\
25&XMMUJ000100.0-250459&00 01 00.0 &$-$25 04 59 &0.851  &BL AGN$^{b}$\\
26&XMMUJ000102.5-245847&00 01 02.5 &$-$24 58 47 &0.433  &BL AGN$^{b}$\\
27&XMMUJ000106.8-250845&00 01 06.8 &$-$25 08 45 &$-$&$-$\\
\multicolumn{6}{c}{MS 0737.9+744 field}\\
28&1E0737.0+7436&07 43 12.5 &$+$74 29 35 &0.332 &BL AGN$^{c}$\\
29&XMMUJ074350.5+743839&07 43 50.5 &$+$74 38 39 &$-$&No cl.\\
30&1SAX J0741.9+7427&07 42 02.2 &$+$74 26 24 &$-$&No cl.\\
31&XMMUJ074351.5+744257&07 43 51.5 &$+$74 42 57 &$-$&No cl. \\
32&XMMUJ074401.5+743041&07 44 01.5 &$+$74 30 41 &$-$&No cl. \\
\multicolumn{6}{c}{Markarian 205 field}\\
33&MS1219.9+7542&12 22 06.6 &$+$75 26 14 &0.238 &NELG$^{d}$\\
34&MS1218.6+7522&12 20 52.0 &$+$75 05 29 &0.646 &BL AGN$^{d}$ \\
35&XMMUJ122258.3+751934&12 22 58.3&$+$75 19 34 &0.257 &NELG$^{d}$ \\
36&XMMUJ122351.3+752224&12 23 51.3&$+$75 22 24 &0.565 &BL AGN$^{d}$\\
37&NGC4291&12 20 15.9 &$+$75 22 09 &0.0058&Galaxy$^{d}$\\
\multicolumn{6}{c}{Abell 1835 field}\\
38&XMMUJ140127.7+025603&14 01 27.7 &$+$02 56 03 &0.265 & BL AGN$^{b}$ \\
39&XMMUJ140053.0+030103&14 00 53.0 &$+$03 01 03 &0.573&BL AGN$^{b}$\\
40&XMMUJ140130.7+024529&14 01 30.7 &$+$02 45 29 &$-$ &No cl.\\
41&XMMUJ140145.0+025330&14 01 45.0 &$+$02 53 30 &$-$$^{\dagger}$ &Gal.$^{b,\dagger}$\\
\hline
\end{tabular}
\end{center}
 Optical classifications and redshifts are from: $^{(a)}$ Fiore et al. (2000), $^{(b)}$ Fiore et al. 2002 (F02; in prep.), $^{(c)}$ Wei et al. (1999) 
and $^{(d)}$ AXIS (Barcons et al. 2001).
$^{(\dagger)}$ There are two possible candidates for the identification of this source: an elliptical galaxy at $z$=0.251 or
an elliptical galaxy at $z$=0.254 (F02).
\end{table*}

\subsection{Spectral analysis}

All spectra are extracted in the 0.3-10 keV band where {\it EPIC} is best calibrated.
The corresponding background regions are extracted from offset positions close to the sources,
in source-free regions and with the same radii of the source regions (i.e. 35 arcsec). For a few cases, 
the extraction radius was reduced due to the presence of another nearby source or of the CCD gaps.
The latest response matrices (released in July 2001) for the corresponding filters are used for the analysis.
All our spectra are also corrected for vignetting. It is worth noting, however, that for off-axis angles $>$10 arcmin the vignetting 
function is energy independent below $\sim$5 keV for all {\it EPIC} detectors (with
 a ratio between off-axis and on-axis counts of $\sim$0.6, Gondoin et al. 2000): since  $\sim$80\%
of our sources (32 out of 41) are detected within 10 arcmin off-axis the spectral correction for vignetting is only significant
for a small number of sources.
Therefore we expect little or almost negligible 
 contamination in the determination of the spectral parameters during our analysis. 
For each source of the sample, we perform a combined spectral fitting using the data of both {\it MOS} and {\it PN} detectors, 
whenever both  datasets are available.
Before spectral fitting, all spectra are binned with a 
minimum of 20 counts per bin in order to be able to apply the $\chi^{2}$ minimization technique. 
In the cases of relatively poor statistics (i.e $<$400 final counts in the 0.3-10 keV band) the minimun number of counts per bin is set to 15
and we apply the Gehrels weighting method (Gehrels 1986) in our analysis. 
However heavy binning of data with poor statistics can result in a loss of spectral 
information, so for a worthwhile comparison we also use the Cash statistic (Cash 1979) to estimate the fit parameters and their 
errors in the {\it PN} data analysis of those sources characterized by only few photons. 
This is a maximum likelihood method which allows the use of unbinned data but it assumes that the counts in a given
channel follow a Poisson distribution, so that it cannot be applied to background-subtracted data.
Hence before performing spectral modelling we accurately parameterize the background in a large source-free region with a model 
consisting of a broken power law plus gaussian lines as suggested in Lumb et al. (2002).
Then we include this model for the background (fixing the values of the parameters and rescaling
the normalizations to the area of the extraction region of the source) in the source spectral parameterization.
However the use of the Cash statistic has a major disadvantage in that it
does not provide a goodness-of-fit {\bf criterion} for comparing different models: therefore in order to establish the best fit model for a source
we apply the $\chi^{2}$ statistic with the Gehrels weighting method and the $F$-test.
The spectral analysis is carried out using {\small XSPEC} v11.0.
In computing fluxes and luminosities of our sources, we apply an encircled energy correction factor in order
to take into account the {\it EPIC} energy encircled fraction (Ghizzardi 2001). 
All luminosities are calculated with $H_{0}$=50 km s$^{-1}$ Mpc$^{-1}$ and $q_{0}$=0.
Throughout the paper, errors are given at the 90\% confidence level for one interesting parameter 
($\Delta\chi^{2}$=2.71; Avni 1976).

\section{Results}
\subsection{Optical and radio properties of the sample}
\begin{figure}[htb]
\begin{center}
\psfig{figure=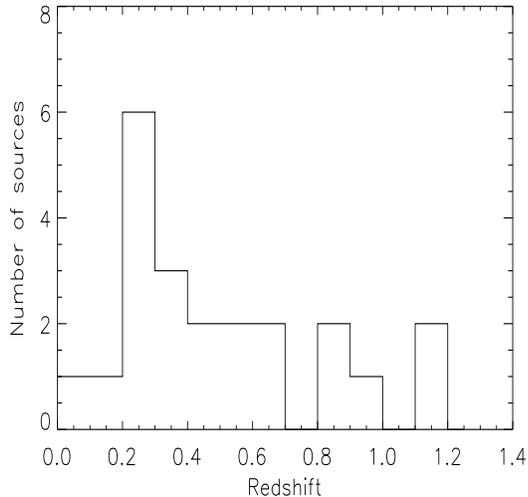,height=7.5cm,width=8cm,angle=0}
\caption{Redshift distribution of the 21 optically identified sources in the sample. The plotted values range from 0.0058 to 1.187. The average redshift
 is $\langle$$z$$\rangle\sim$0.5.}
\label{fig:redshift}
\end{center}
\end{figure}

As reported in Table 2, 21 of our 41 X-ray sources ($\sim$50\% of the sample) are already optically identified: 9 from the literature,
 4 from the {\it XMM-Newton} International Survey program ({\it AXIS}, Barcons et al. 2001) and 10 from Fiore 
et al. (in preparation, hereafter F02). 
Of these, the majority (15) are broad line AGNs,
2 are narrow emission line galaxies (NELGs), 1 is a Seyfert 2 galaxy, 2 are normal galaxies and one is a red quasar.
Source n.41 has two possible optical counterparts, both of which are normal elliptical galaxies at $z$=0.251 and $z$=0.254,
 respectively (F02).
The redshift distribution of these optically identified sources is shown in Fig.~\ref{fig:redshift}.

Source n.3 (CXOU J031238.9$-$765134) was first discovered by {\it Chandra} and identified 
with a bulge dominated normal galaxy without strong optical emission lines at $z$=0.158 (called 'P3' in Fiore et al. 2000).
Combined with its high X-ray luminosity, this makes it a classic example of an
'optically dull X-ray loud' galaxy (Elvis et al. 1981).
In the optical image the galaxy appears to be extended with the {\it Chandra} source centered on its nucleus.  
Source n.33 (MS1219.9$+$7542) was detected for the first time by the {\it Einstein} observatory (Gioia et al. 1990), and
was first optically identified with a cluster of galaxies by Stocke et al. (1991) in the {\it EMSS} survey.
Subsequently, Rector et al. (2000) classified MS 1219.9+7542 as a BL Lac object on the basis of its spectral energy distribution (SED).
Very recently, however, Barcons et al. (2001) unambiguously classified MS1219.9$+$7542 as a narrow emission line galaxy (NELG) on 
the basis of an observation performed with the 2.5m INT telescope.
Source n.37 (NGC 4291) is an early-type (E2) absorption line galaxy at redshift $z$=0.0058 (Ho, Filippenko \& Sargent 1997),
which is also weakly detected in a 8.5 GHz NRAO VLA observation ($P_{8.5GHz}<$6.3 $\times$  10$^{18}$ W Hz$^{-1}$, Wrobel \& Herrnstein 2000).
Stellar dynamical measurements indicate that NGC 4291 has a massive central black hole of 
1.8 $\times$ 10$^{8}$ solar masses (Gebhardt et al. 2000). This nearby galaxy appears extended 
in the {\it EPIC} image. 
The centroid of the hard X-ray source ($\alpha$(J2000)=12$^{h}$20$^{m}$15$^{s}$, $\delta$(J2000)=75$^{\circ}$22$^{\prime}$09$^{\prime\prime}$) is displaced by $\sim$9 arcsec from the optical
center of the galaxy ($\alpha$(J2000)=12$^{h}$20$^{m}$17$^{s}$, $\delta$(J2000)=
75$^{\circ}$22$^{\prime}$18$^{\prime\prime}$). This value is larger than the {\it XMM}
 astrometric errorbox (6 arcsec): 
 the X-ray source is however well inside the optical contour of NGC 4291. 
We choose a 30 arcsec extraction radius 
in order to avoid contamination from the nearby X-ray
source 1AX J122024+7521 (Ueda et al. 2001) and the previously unknown X-ray source XMMUJ122005.1+752143,
which are at 55 and 49 arcsec from NGC 4291, respectively.
 
Finally, the sample is cross-correlated with the FIRST (Becker, White \& Helfand 1995) and the NVSS (Condon et al. 1998) 
radio catalogs in order to 
identify any possible radio counterparts.  Only 2 out of 7 fields analyzed here have been covered by the FIRST survey 
while the NVSS covers all but one fields (PKS0312-770) but with a higher flux limit.  
We find that only three X-ray sources have a radio counterpart: sources n. 23, 25 and 
38 coincide (within a few arcsec) with the radio sources NVSS000122.7-250018.6 ($S_{1.4GHz}$=69.2 mJy), 
NVSSJ000100-250503 ($S_{1.4GHz}$=130mJy) and FIRST 140127.59+025606.8 ($S_{1.4GHz}$=1.54 mJy), respectively. 
All these sources have optical-to-radio indices $\alpha_{ro}$ $\geq$ 0.3 and can therefore be classified as 
radio loud AGNs.

\begin{figure}[h!]
\begin{center}
\psfig{figure=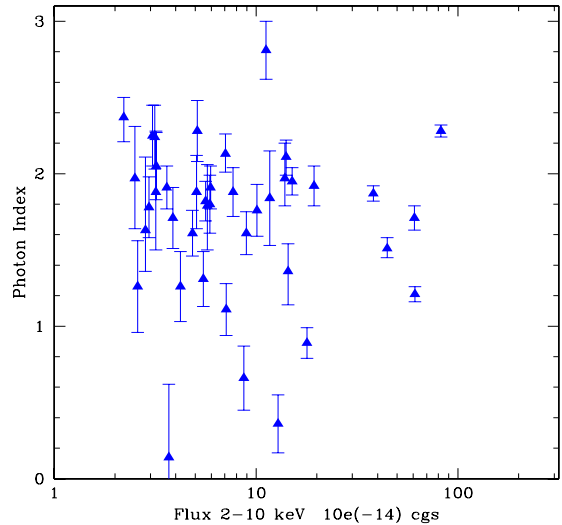,height=7.5cm,width=8cm,angle=0}
\caption{Broad-band (0.3-10 keV) photon {\bf indices} obtained using a power law plus Galactic absorption (SPL) model 
 versus 2-10 keV flux for each source in our sample.}
\label{fig:Gamma_F_modA}
\end{center}
\end{figure} 

\begin{table*}[htb]
\caption{Fits with a single power law model with Galactic absorption (model SPL).}
\label{tab3}
\begin{center}
\begin{tabular}{cccccc}
\hline
\multicolumn{1}{c} {N} &
\multicolumn{1}{c} {Source name} &
\multicolumn{1}{c} {$N_{\rm H}^{\rm Gal, (a)}$}&
\multicolumn{1}{c} {$\Gamma^{(b)}$} &
\multicolumn{1}{c} {$\chi^{2}_{\nu}/$(d.o.f.)}&
\multicolumn{1}{c} {Best$-$fit}\\ 
  && (10$^{20}$ cm$^{-2}$)   && &\\
\hline
\hline 
1&CXOU J031015.9$-$765131 &8 &1.87$^{+0.05}_{-0.05}$ &0.90/250&Yes \\
2&CXOU J031209.2$-$765213 &8 &2.24$^{+0.21}_{-0.21}$ &0.91/52&Yes \\		
3&CXOU J031238.9$-$765134 &8 &1.03$^{+0.52}_{-0.52}$/(1.26$^{+0.30}_{-0.30}$)& 0.69/11&No \\		
4&CXOU J031253.8$-$765415 &8 &1.61$^{+0.14}_{-0.14}$ &1.02/39&Yes \\		
5&CXOU J031312.1$-$765431 &8 &1.88$^{+0.16}_{-0.16}$ &1.22/48&Yes \\	
6&CXOU J031314.5$-$765557 &8 &1.95$^{+0.09}_{-0.09}$ &0.79/81&Yes \\		
7&XMMU J030911.9$-$765824 &8 &0.15$^{+0.30}_{-0.41}$/(0.36$^{+0.19}_{-0.19}$) & 1.28/28&No\\		
8&XMMU J031049.6$-$763901 &8 &1.36$^{+0.18}_{-0.22}$ &1.07/35&Yes\\		
9&XMMU J031105.1$-$765156 &8 &1.32$^{+0.76}_{-0.61}$/(1.88$^{+0.41}_{-0.38}$)&0.81/11&Yes\\		
10&XMMU J123110.6$+$641851 &2 & 1.91$^{+0.14}_{-0.14}$&0.76/37&Yes\\	
11&XMMU J123116.3$+$641114 &2 &1.82$^{+0.13}_{-0.12}$& 1.00/52&Yes\\		
12&XMMU J123218.6$+$640309 &2 &1.88$^{+0.24}_{-0.24}$ & 0.71/24&Yes \\	
13&XMMU J123214.2$+$640459 &2 &1.40$^{+0.30}_{-0.30}$/(1.31$^{+0.18}_{-0.18}$)&0.92/24&Yes\\		
14&XMMU J123013.4$+$642505 &2 &1.79$^{+0.27}_{-0.29}$ & 1.14/22&Yes\\		 
15&XMMU J123049.9$+$640845 &2 &2.37$^{+0.13}_{-0.16}$ & 1.05/31&Yes\\		
16&XMMU J123058.5$+$641726 &2 &1.08$^{+0.34}_{-0.34}$/(1.26$^{+0.23}_{-0.23}$)&0.32/9&Yes\\
17&XMMU J133730.8$+$242305&1.2 &2.25$^{+0.21}_{-0.21}$&0.58/24&Yes\\	
18&XMMU J133649.3$+$242004&1.2 &2.05$^{+0.22}_{-0.22}$ &0.94/24&Yes\\		
19&XMMU J133807.4$+$242411&1.2 &2.13$^{+0.11}_{-0.12}$ &1.21/72&No\\ 		
20&XMMU J133747.4$+$242728&1.2 &1.92$^{+0.34}_{-0.33}$ &0.80/24&Yes\\		
21&XMMU J133712.6$+$243252&1.2 &1.66$^{+0.43}_{-0.28}$&1.46/17&Yes\\		
22&XMMU J000031.7$-$255459&2 &	2.13$^{+0.13}_{-0.12}$&0.91/19 &Yes\\		
23&XMMU J000122.8$-$250019&2 &	1.97$^{+0.23}_{-0.18}$&1.11/31&Yes \\		
24&XMMU J000027.7$-$250441&2 &	1.91$^{+0.14}_{-0.14}$&0.99/46&Yes\\		
25&XMMU J000100.0$-$250459&2 &	0.89$^{+0.10}_{-0.10}$&1.44/28&No\\	
26&XMMU J000102.5$-$245847&2 &	2.28$^{+0.20}_{-0.20}$&0.85/52&Yes\\
27&XMMU J000106.8$-$250845&2 &  1.61$^{+0.39}_{-0.33}$/(1.71$^{+0.20}_{-0.20}$)&0.64/19&Yes\\		
28&1E0737.0$+$7436       &3.5 &	2.28$^{+0.04}_{-0.04}$  &1.05/227&No\\	
29&XMMUJ074350.5$+$743839&3.5 &1.08$^{+0.25}_{-0.25}$/(1.11$^{+0.17}_{-0.17}$)&1.11/13 &No\\	
30&1SAX J0741.9+7427      &3.5&	1.92$^{+0.13}_{-0.13}$  &1.11/64&Yes\\		
31&XMMUJ074351.5$+$744257&3.5 &	1.84$^{+0.31}_{-0.31}$  &0.97/29&Yes\\	
32&XMMUJ074401.5$+$743041&3.5 &1.71$^{+0.34}_{-0.30}$/(1.78$^{+0.20}_{-0.20}$)&0.59/11&Yes\\
33&MS1219.9$+$7542         &3&1.51$^{+0.07}_{-0.06}$&0.86/87&No\\
34&MS1218.6$+$7522         &3&1.71$^{+0.08}_{-0.08}$&0.69/100&Yes\\	
35&XMMU J122258.3$+$751934 &3&1.41$^{+0.27}_{-0.33}$/(1.61$^{+0.15}_{-0.15}$)&0.93/18&Yes\\
36&XMMU J122351.3$+$752224 &3&1.76$^{+0.17}_{-0.17}$&0.84/31&Yes\\
37&NGC4291                 &3&2.81$^{+0.19}_{-0.19}$&5.26/72&No\\	
38&XMMU J140127.7$+$025603 &2.3&	1.21$^{+0.05}_{-0.05}$&1.55/215&No\\	
39&XMMU J140053.0$+$030103 &2.3&	1.80$^{+0.19}_{-0.19}$&0.90/37&Yes\\	
40&XMMU J140130.7$+$024529 &2.3&	0.19$^{+0.52}_{-0.62}$/(0.14$^{+0.48}_{-0.28}$)& 0.87/7&No\\			
41&XMMU J140145.0$+$025330 &2.3&         0.76$^{+0.35}_{-0.35}$/(0.66$^{+0.21}_{-0.21}$)& 0.85/19&Yes\\
\hline
\end{tabular}
\end{center}
$^{(a)}$Galactic column density are taken from Stark et al. (1992).  $^{(b)}$ The values reported in brackets are obtained
using the C statistic (Cash 1979), see text for details.
\end{table*}

\begin{table*}[htb]
\caption{Fits with a single power law model with extra absorption component (model APL)}
\label{tab_abs}
\begin{center}
\begin{tabular}{cccccc}
\hline
\multicolumn{1}{c} {Source n.} &
\multicolumn{1}{c} {$\Gamma^{(a)}$} &
\multicolumn{1}{c} {$N_{\rm H}^{(a)}$}&
\multicolumn{1}{c} {$F-$statistic} &
\multicolumn{1}{c} {C.l.}&
\multicolumn{1}{c} {Best-fit}\\
&&(10$^{21}$ cm$^{-2}$)&&&\\
\hline\hline\\ 
3&1.84$^{+1.95}_{-1.10}$/(2.12$^{+0.68}_{-0.24}$)&7.1$^{+18.0}_{-7.1}$/(11.8$^{+5.7}_{-5.7}$)&3.1&90.0\%&Yes\\
7&1.20$^{+1.26}_{-1.05}$/(1.39$^{+0.53}_{-0.53}$)&48.5$^{+70.3}_{-48.5}$/(37.6$^{+26.8}_{-18.5}$)&1.14&80.0\%&No\\
25&1.31$^{+0.20}_{-0.19}$&5.2$^{+3.0}_{-2.2}$&18.0&99.99\%&Yes\\
29&1.75$^{+0.72}_{-0.47}$/(1.59$^{+0.30}_{-0.30}$)&2.7$^{+2.8}_{-1.8}$/(2.1$^{+1.8}_{-0.9}$)&7.9&98.5\%&Yes\\
33&1.64$^{+0.09}_{-0.09}$&0.59$^{+0.23}_{-0.23}$&4.8 &97.0\%&Yes\\
37&$>$8.69&10.7$^{+0.5}_{-1.7}$&65.1&$>$99.99\%&No\\
38&1.66$^{+0.09}_{-0.09}$&1.5$^{+0.2}_{-0.2}$&98&$>$99.99\%&No\\
40&2.15$^{+3.37}_{-1.84}$/(1.74$^{+2.75}_{-0.47}$)&23.5$^{+120.8}_{-18.5}$/(18.1$^{+10.4}_{-5.90}$)&6.1&96.0\%&Yes\\
\hline
\end{tabular} 
\end{center}
$^{(a)}$ The values reported in brackets are obtained
using the C statistics (Cash 1979), see text for details.
\end{table*}

\subsection{X-ray spectral properties}
\subsubsection{Single power law model (SPL)}
In order to understand the shape of the continuum, the spectrum of each source is first fitted  
with a simple model consisting of a power law with absorption 
fixed at the Galactic value (hereafter, model SPL).
The results of these fits are reported in Table 3 and graphically displayed in Fig.~\ref{fig:Gamma_F_modA}
where we plot $\Gamma$ versus 2-10 keV flux. 
The resulting photon indices range from 0.15 to 2.8, 
with the majority of the sources clustering at around 1.9, i.e. close to the canonical value for
broad line unabsorbed AGNs.  One source in Fig.~\ref{fig:Gamma_F_modA} is compatible 
with an inverted (negative) spectral slope and
another has $\Gamma$$\sim$3. These extreme cases will be discussed in detail in the next Sections. 
The results obtained by applying the Cash statistic are fully consistent with those obtained  using
the $\chi^{2}$ (see Table 3).
We also check out the possible presence of systematic variations (i.e. hardening or softening) of the photon index with the off-axis
 distance of the sources but we find no particular trend. 
Possible artifacts of the PSF emerging at large radii 
have been also rejected by Bocchino et al. (2001) after a detailed 
 {\it MOS} and {\it PN} spectral analysis of 3C58. 
Interestingly enough, it appears from Fig.~\ref{fig:Gamma_F_modA} that very flat $\Gamma$ ($\leq$1.1-1.2)
 are present at different flux levels and no trend of $\Gamma$ versus flux is evident in the data.  

\subsubsection{Absorbed power law model (APL)}
\label{apl_s}
 A flat photon index could indicate that  
the simple power law model is not appropriate. 
In fact, typically flat X-ray spectra are obtained when absorption intrinsic to the source is present. 
We have therefore refitted our data using a power law but absorbed this time both by Galactic and intrinsic absorption (model APL).
Table 4 lists all sources (8 out of 41) for which there is evidence of excess 
absorption on the basis of a $\chi^{2}$ improvement in the fit.
In source n.3 the fit improvement is not so significant (at $>$90\%) but the slope is steeper ($\Gamma$$\sim$1.84) 
than without absorption ($\Gamma$$\sim$1.03). 
In source n.7, the fit improves only slightly but this source clearly requires an additional soft component 
(see Section~\ref{complex}).
 In all other sources (n.25,
 29, 33, 37, 38 and 41) the improvement is highly significant (at $>$97\%). 

In any case it is evident from Table 4 that the power laws steepen to more standard $\Gamma$ values in most objects, 
while the column densities obtained range from $\sim$6 to $\sim$200 $\times$ 10$^{20}$ cm$^{-2}$, 
i.e. a very broad range of values: in source n.37 $\Gamma$ is intrinsically steep but this is due to the peculiar
spectral shape of this source (see Section~\ref{complex}).

As expected the only known Seyfert 2 (n.7) present in the sample is the source with the largest value of $N_{\rm H}$. 
Note however that other objects characterized by narrow lines (n.4, n.33, n.35), 
and so potentially obscured, show very low column density values. 
Furthermore, although not yet optically classified, XMMU J140130.7$+$024529 (n.40) is very likely associated with a type 2
object just on the basis of the column density result obtained here.
Significant absorption is also found in two broad line 
AGNs (n.25 and n.38), indicating that also this type of object can contain large amounts of gas; 
furthermore 2 out of 3 radio loud sources in our sample are obscured in X-rays.
 Even more interesting is the presence of absorption in two out of three normal galaxies of the sample.
Under the assumption that the dust to gas ratio in these objects is similar to our own Galaxy, 
the observed column densities imply an optical extinction A$_{\rm V}$$\sim$1-2 mag, 
not high enough to obscure the broad line region of an AGN, and so confirming their optical classification
 as normal galaxies. 
Finally source n.41 does not require any additional absorption component and its spectrum
appears to be intrinsically flat ($\Gamma$$\sim$0.7). We have also tried imposing a $\Gamma$=1.9, resulting in
a column density $N_{\rm H}$=1.1$^{+1.2}_{-0.7}$ $\times$ 10$^{22}$ cm$^{-2}$ but this fit is worse than that found with the SPL model.
However the lack of optical identification for about half of the sources in the sample and  
the poor statistics characterizing some faint sources could lead to an artificial underestimate 
of $N_{\rm H}$.
In fact, it is worth noting that some sources with flat best-fit slopes ($\Gamma<$1.5) 
could be obscured by large absorption columns but  the low statistics may  prevent us 
from making a correct measurement of the true absorption column density.
We have faced both these problems by 
imposing a photon index of $\Gamma$=1.9 to all sources
and $z$=1 to optically unidentified objects, before re-estimating the $N_{\rm H}$ values.
Results of this spectral fitting will be discussed in Section~\ref{comp_models}.

\subsubsection{More complex models}
\label{complex}

\begin{figure}[htb]
\psfig{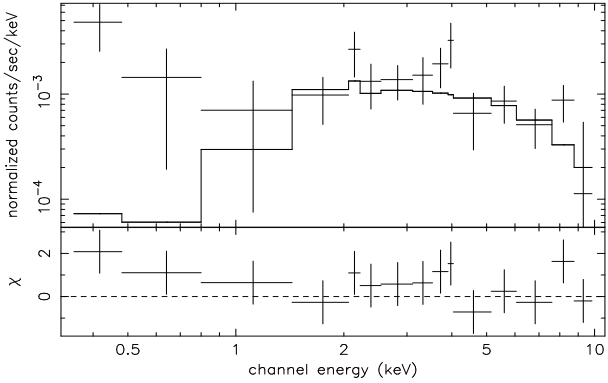}
\caption{The {\it PN} spectrum of XMMU J030911.9-765824 (n.7) ({\it upper panel}) fitted with a power law modified by 
an extra X-ray absorption component (model APL, see Table 4). The residuals of the fit are also shown ({\it lower panel}). 
Note the excess in the soft portion of the spectrum.}
\label{fig:sp_0309-7658}
\end{figure}

\begin{figure}[htb]
\psfig{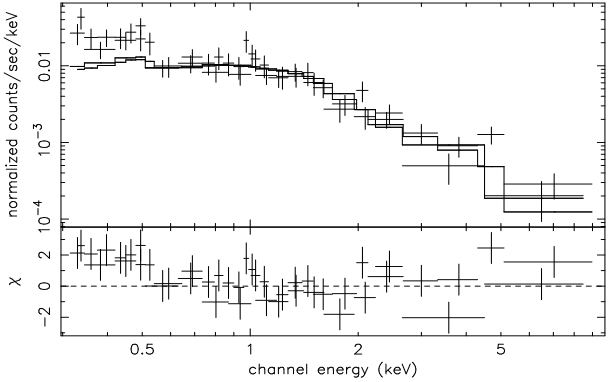}
\caption{{\it MOS1}+{\it MOS2} spectra of the unidentified source XMMU J133807+242411 (n.19) fitted by a simple power law plus Galactic absorption model ({\it upper panel}). The residuals ({\it lower panel}) show evidence of a soft excess emission component.}
\label{fig:irasC}
\end{figure}

\begin{figure}[htb]
\psfig{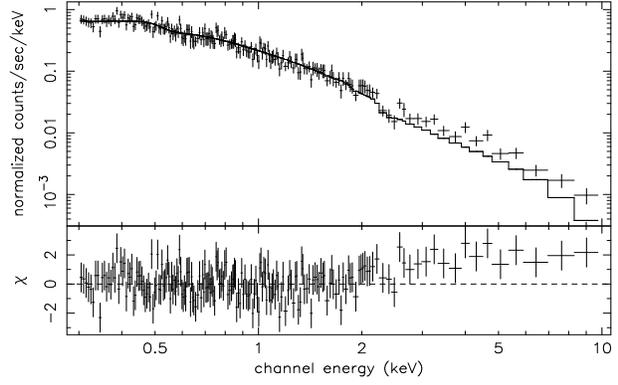}
\caption{The $PN$ spectrum  ({\it upper panel}) and residuals ({\it lower panel}) of the broad line
 QSO 1E 0737.0+7436 (n.28) when fitted by a simple power law plus Galactic absorption (SPL) model (Table 3).}
\label{fig:1e07_res}
\end{figure}

\begin{figure}[htb]
\psfig{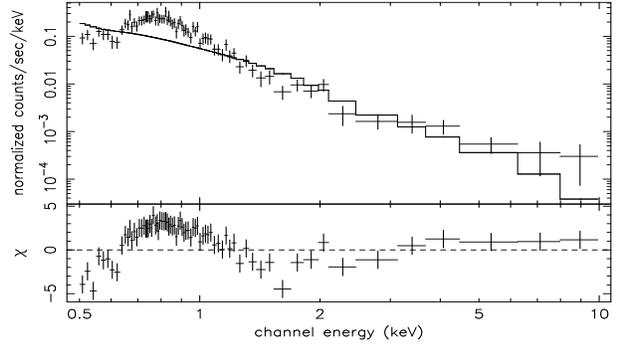}
\caption{The  {\it EPIC PN} spectrum of NGC 4291 (n.37) fitted by a power law ({\it upper panel}) and relative residuals
 ({\it lower panel}). The large residuals indicate that this model is inadequate to describe the present data (Table 3).}
\label{fig:ngc4291_sp_pl}
\end{figure}

After the introduction of models SPL and APL four sources (n.7, n.19, n.28, n.37) are still not satisfactorily fitted; 
in fact the residuals  show (see Fig.~\ref{fig:sp_0309-7658}, \ref{fig:irasC}, \ref{fig:1e07_res}, \ref{fig:ngc4291_sp_pl})
 that the presence of a soft excess is these objects is evident. 
We have therefore refitted these sources introducing an extra spectral component in the form
of a power law (PL in Table 4) or a thermal model (TM in Table 4) in addition to the primary power law continuum. 
The photon index of the secondary power law is left 
free to vary in all sources whereas in source n.7 it is fixed to the value 
of the primary power law photon index as expected in a type 2 object where the soft excess is often due to a scattered
component (Turner et al. 1997). The thermal model is parameterized by Raymond-Smith plasma.

 Source n.19 is not optically classified  and the introduction of a thermal component is dictated by 
the high value of the secondary power law photon index ($\Gamma$=5.17$^{+1.64}_{-0.43}$). 
In source  n.28, a double power law model is a good fit to the data and the change in slope
 is located at 1.30$^{+0.31}_{-0.20}$ keV.
Many radio-quiet QSOs, as source n.28 has been optically classified, show such an excess in
the soft X-ray band (Reeves \& Turner 2000, George et al. 2000, Mineo et al. 2000).
This excess is likely due to the high energy tail of the UV bump 
and is often parameterized by a blackbody component.
In fact using such a component in place of the second power law gives an equally good fit 
(at $>$99\% confidence level with respect to model SPL) and provides a $\Gamma$=1.96$^{+0.10}_{-0.09}$ 
 and a $k$T=0.13$^{+0.02}_{-0.02}$ keV in the quasar rest frame.

\begin{table*}[htb]
\caption{Fits with an absorbed/unabsorbed power law plus soft thermal(TM)/power law (PL) component model.}
\label{tab_soft}
\begin{center}
\begin{tabular}{cccccccc}
\hline\\
\multicolumn{1}{c} {Source n.} &
\multicolumn{1}{c} {Model} &
\multicolumn{1}{c} {$\Gamma_{soft}/k$T}&
\multicolumn{1}{c} {$N_{\rm H}$}&
\multicolumn{1}{c} {$\Gamma_{hard}$}&
\multicolumn{1}{c} {$F-$statistic} &
\multicolumn{1}{c} {C.l.}& 
\multicolumn{1}{c} {Best-fit}\\
&&(eV)&(10$^{21}$ cm$^{-2}$)&&&&\\
\hline\hline\\
7&PL&$\equiv\Gamma_{hard}$&94$^{+92}_{-92}$/(98$^{+41}_{-33}$)$^{(a)}$&1.76$^{+0.50}_{-1.17}$/(2.17$^{+0.48}_{-0.29}$)$^{(a)}$&4.4&96\%&Yes\\
19&TM&150$^{+50}_{-110}$&$\equiv$$N^{\rm Gal}_{\rm H}$&1.74$^{+0.23}_{-0.19}$&8.1&$>$99\%&Yes\\
28&PL&2.47$^{+0.08}_{-0.07}$&$\equiv$$N^{\rm Gal}_{\rm H}$&1.93$^{+0.10}_{-0.10}$&23.5&$>99$\%&Yes\\
37&TM&360$^{+50}_{-20}$&2.3$^{+1.1}_{-0.8}$&1.11$^{+0.50}_{-0.48}$&96.7&$>$99.99\%&Yes\\
\hline
\end{tabular} 
\end{center}
$^{(a)}$ The values reported in brackets are obtained
using the C statistic (Cash 1979), see text for details.
\end{table*}

\begin{figure}[htb]
\psfig{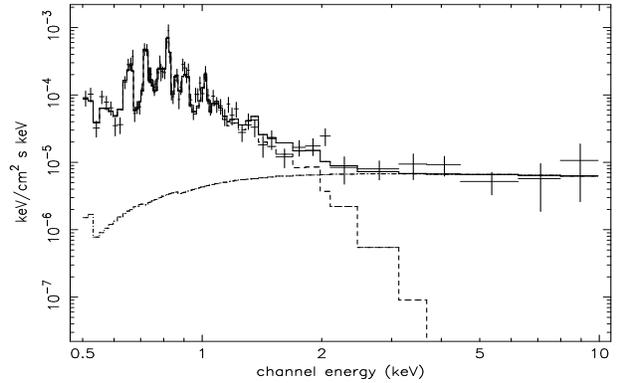}
\caption{The {\it EPIC PN} $\nu$F$\nu$ spectrum of NGC 4291. We also show the two components of the best fit model 
(see Table 6).}
\label{fig:ngc4291sp_all}
\end{figure}
\begin{figure}[htb]
\psfig{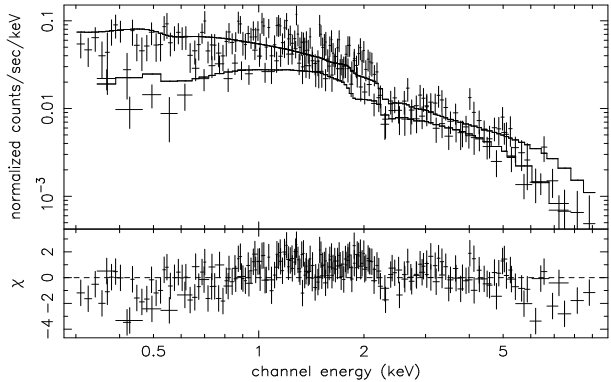}
\caption{The {\it PN}+{\it MOS2} spectrum of {\it XMMU J140127.7+025603.4} fitted with a simple
power law (model SPL). Absorption features in the soft as well as in the hard portion of the spectrum are 
evident.}
\label{fig:x14_modA}
\end{figure}
Source n.37 (NGC4291) is a normal galaxy and so a thermal component is expected:
for the X-ray emitting gas we infer a rather low temperature ($k$T$\sim$0.4 keV) and metallicity $Z/Z_{\odot}$=0.07$^{+0.04}_{-0.02}$.
The introduction of the soft component solves the problem of the rather steep spectral slope but reveals the presence
 of a hard tail in this object (e.g. Fig.~\ref{fig:ngc4291sp_all}). 
A similar feature has also been found in other X-ray early-type 
galaxies (Allen et al. 2000, Matsumoto et al. 1997) and is possibly 
due either to the presence of an obscured nucleus or to the summed contributions of discrete X-ray sources in the galaxy. 
If we impose a slope of 1.9 to the power law  component likely associated with the nucleus,  
we obtain a column density of 41$^{+30}_{-21}$ $\times$ 10$^{21}$ cm$^{-2}$, thus invoking an obscured nucleus
could be a viable explanation for the hard tail. However, using this model  we find 
{\it L$_{\rm 2-10}$}=1.7 $\times$ 10$^{40}$ erg s$^{-1}$ which provides an  $L_{X}/L_{Eddington}$ ratio $\sim$4 $\times$ 10$^{-6}$,
 too low for a normal AGN.  Recent {\it Chandra} results 
(Sarazin, Irwing \& Bregman 2001, Loewenstein et al. 2001) indicate that the  bulk of the 2-10 keV emission in a
number of early-type galaxies originates from discrete galactic sources like LMXBs, HMXBs and/or galactic
 BHs (Fabbiano, Zezas \& Murray 2001; Foschini et al. 2002) and so it is likely that this contribution 
is present in NGC4291 too. 
Unfortunately we cannot investigate more deeply 
the origin of the hard  X-ray tail because of the limited  spatial resolution of {\it EPIC} with respect to
 {\it Chandra} and the low signal to noise ratio at high energies.
An observation with higher angular resolution (i.e. with {\it Chandra}) and longer exposure could 
instead place stringent limits on
the relative contribution of the diffuse gas, discrete sources and a possible hidden low luminosity 
active nucleus to the total 2-10 keV luminosity of this source.

Finally source n.38, a broad line radio loud quasar, is characterized by a warm absorber 
as shown in Fig.~\ref{fig:x14_modA}. 
In the soft portion of the spectrum we find an edge at 0.82$\pm$0.10 keV (quasar-frame) with an optical depth 
$\tau$=0.42$^{+0.15}_{-0.11}$, likely due to OVIII $K-$shell absorption.
With this additional component, whose significance is at $\sim$95\% level, the associated \xnu~is equal to 218(212).
The other edge is found at $E_{\rm abs}$=7.12$^{+0.52}_{-0.43}$ keV (quasar-frame) with an optical depth 
$\tau$=0.71$^{+0.46}_{-0.38}$ and a significance  $>$95\% confidence level ($F$-test value =4.5). 
This edge is consistent with FeI-XVIII $K-$shell at 90\% confidence for $\tau$$>$0.3, assuming that the absorbing material
is intrinsic to the source. Including this feature in the spectral fitting gives a 
slightly harder photon index ($\Gamma$=1.55$^{+0.05}_{-0.05}$).
 
We have also checked in the optically identified sources the possible presence 
of an emission feature centered at 6.4 keV in the source rest frame.
No convincing evidence for this line has been found, and in all of the sources considered the upper limits on EW we can set are
compatible with the values usually observed in active galaxies ($\leq$350 eV).

\begin{table*}[]
\caption{{\it XMM-Newton} properties of the sample. Fluxes and luminosities are obtained using the best fit model listed in column 2
for each source of the sample.}
\label{tab_flux}
\begin{center}
\begin{tabular}{ccccc}
\hline
\multicolumn{1}{c} {Source n.} &
\multicolumn{1}{c} {Best fit } &
\multicolumn{1}{c} {$F_{0.5-2}$} &
\multicolumn{1}{c} {$F_{2-10}$}&
\multicolumn{1}{c} {$L_{2-10}$}\\
&Model$^{\dagger}$&(10$^{-14}$ \cgs)&(10$^{-14}$ \cgs)&(10$^{44}$ erg s$^{-1}$)\\
\hline\hline\\
1	&SPL&	22.10&	38.1&	52.7	\\
2	&SPL&	4.50&	3.15&	3.2	\\
3	&APL&	0.66&	 2.59&	3.5$\times$10$^{-2}$\\
4	&SPL&	3.41&	8.95&	2.5	\\
5	&SPL&	4.31&	7.70&	9.1	\\
6	&SPL&	10.13&	15.14&	1.7	\\
7	&PL&	0.55&12.85&	 6.8	\\
8	&SPL&	6.25&	14.42&	1.3	\\
9	&SPL&	0.83&	3.19&	$-$	\\
10	&SPL&	2.62&	3.61	&	$-$	\\  
11	&SPL&	 3.60&	 5.62	&	$-$   \\
12	&SPL&	3.52&	5.07	&	$-$	\\  
13	&SPL&	 1.84&	5.48	&	$-$	\\  
14	&SPL&	3.51&	5.74	&	$-$\\	
15	&SPL&	3.95&	2.21	&	$-$	\\  
16	&SPL& 	 0.88& 4.22   &$-$	\\  
17	&SPL&3.86&	3.07	&	$-$	\\	 
18	&SPL&2.61&	3.20	&	$-$	\\	 
19	&TM&8.52&	14.10	&	 $-$	\\ 
20	&SPL&1.88&	2.51	&	 $-$	\\ 
21	&SPL& 1.43& 2.83	&	 $-$	\\
22	&SPL&7.11&	7.06	&0.3\\
23	&SPL&9.08&	13.88	&12.2	\\
24	&SPL&4.37&	5.95	&0.4\\
25	&APL&4.10&	17.88	&7.3	\\
26	&SPL&6.40&	5.12	&0.7\\
27	&SPL&2.10&	 3.87   &$-$	\\	
28	&PL&72.32	&82.42	&5.3	\\
29	&APL& 2.47	&7.11	&$-$	\\
30	&SPL&14.41	&19.38	&$-$	\\
31	&SPL&7.50	&11.70	&$-$	\\
32	&SPL&1.74      & 2.95   &$-$	\\
33	&APL&		18.33&	44.6	&1.3	\\
34	&SPL&	30.62&	60.93	&15.9	\\
35	&SPL&	 1.96&	 4.84&0.2\\
36	&SPL&	6.76&	10.11	&2.1	\\
37	&TM&	24.41&   11.21	&1.7$\times$10$^{-4}$\\
38	&WA&	22.23	&61.22	&2.2	\\
39	&SPL&	3.62	&5.91	&1.3	\\
40	&APL&		0.31	& 3.7&$-$	\\	
41	&SPL&		 1.10	& 8.7&$-$\\
\hline\\
\end{tabular} 
\end{center}
$^{\dagger}$SPL$=$power law model + Galactic absorption; ABL$=$SPL + additional absorption; TM$=$SPL (or APL) + thermal model 
for the soft excess component; PL$=$SPL (or APL) + non-thermal model for the soft excess component; WA$=$SPL (or APL) + warm 
absorber features.
\end{table*}
\subsubsection{Summary}
About 75\% of our sources have not been detected in X-rays before and except for 6 objects (n.1 to n.6 in Table 2) 
in the PKS0312-770 sky field (for which we find similar results to Lumb, Guainazzi \& Gondoin 2001), all others have not 
been spectroscopically studied above 2 keV. 
31 out of 41 sources are well fitted with a simple power law plus Galactic absorption (SPL) model (see Table 3).
In 7 sources an extra absorption component is required at $>$95\% confidence level (see Table 4  and 5).
Furthermore another source (n.3 in Table 4) shows a very flat photon index which is indicative of obscuration, 
but an accurate spectral fitting 
is limited by the poor statistics.
In four sources we find significant soft excess emission, which can be parameterized with a thermal (TM) model in two cases 
and with a power law (PL) in another case (source n.7, a Seyfert 2 galaxy) while in one object (n.28), both 
the TM and PL models provide an equally good fit. One source (n.38) is characterized by warm
 absorber features both in the soft and in the hard portion of its 
spectrum (indicated as model WA in Table 6).
 In Table 6 we report the best fit models found for all our of the sources with the corresponding fluxes in the 0.5-2 keV and 
in the 2-10 keV band as well 
as the unabsorbed 2-10 keV luminosity. 
The hard X-ray fluxes range from $\sim$3 to $\sim$80 $\times$ 
10$^{-14}$ \cgs, with more than 60\% of the sources having $F_{2-10}<$10$^{-13}$ \cgs.
The luminosities span from $\sim$2 $\times$ 10$^{40}$ to $\sim$5 $\times$ 10$^{45}$ erg s$^{-1}$. 
Sources with a hard X-ray luminosity $\geq$10$^{42}$ erg s$^{-1}$ match well with their optical classification as AGNs, 
except for the two intriguing 'normal' galaxies (n.3 and n.41) which are too X-ray bright for their class. 

\section{Discussion}
\subsection{Photon index}
\label{slope}
Using model SPL over the 0.3-10 keV band (see Table 3) we find an average spectral index
 $\langle\Gamma\rangle$=1.67$\pm$0.04 (or $\langle\Gamma\rangle$=1.69$\pm$0.04 if we consider only 
the subsample consisting of the 27 sources with a 2-10 keV flux brighter than $\sim$5 $\times$ 10$^{-14}$ \cgs, 
the value above which we are complete).
Using the best fit model (see Table 6) for 40 out of 41 objects\footnote{We exclude NGC4291 (n.37) because its spectrum is 
better described by a thermal model (e.g. Table 6) 
as expected on the basis of its optical classification.},
the  average photon index becomes even steeper ($\langle\Gamma\rangle$=1.74$\pm$0.09).
 This value matches well with those
typical of broad line quasars as found in previous works 
(Lawson \& Turner 1997; Reeves \& Turner 2000) and it is also consistent with the fact
that the large majority of our identified sources are indeed broad line AGNs (see Table 2).
Concerning the subsample of sources with F$_{2-10}\geq$$\sim$5 $\times$ 10$^{-14}$ \cgs,
the average slope we find is fully consistent with the results in Ishisaki et al. (2001) based on the 1-7 keV hardness ratio analysis
of 29 hard X-ray selected sources with fluxes 10$^{-13}$$\leq$$F_{1-7}\leq$10$^{-14}$ \cgs~ found in 
a deep {\it ASCA} exposure of the Lockman Hole field.
Indeed the above authors reported an apparent photon index of $\langle\Gamma\rangle$=1.65$\pm$0.10. 
Our value is instead somewhat steeper than that found using the {\it ASCA LSS} survey 
($\langle\Gamma\rangle$=1.51$\pm$0.05; Ueda et al. 1999) obtained from a stacked spectra analysis of 39 X-ray serendipitous 
sources selected in the 2-10 keV band with $F_{2-10}\geq$8 $\times$ 10$^{-14}$ \cgs.
Ueda et al. (1999) noted however that their result was affected by an artificial spectral flattening 
due to their hard X-ray selection criterion. 

On the other hand, recent {\it Chandra} and {\it XMM-Newton} deep surveys indicate that the
 bulk of sources with flat spectra 
(intrinsic and/or due to obscuration) which are required to solve the CXB spectral paradox, emerges at 2-10 keV
fluxes fainter than 10$^{-14}$ \cgs.
In fact, our value of $\Gamma$$\sim$1.6-1.7 is consistent with the results 
obtained from the stacked spectra analysis of the {\it Chandra} Deep Field South (CDFS), 
where Giacconi et al. (2001) found  $\Gamma$=1.71$\pm$0.07.
A progressive hardening of the average slope towards fainter fluxes which is necessary to explain the 2-10 keV CXB spectral shape,
has however been measured by Tozzi et al. (2001).
Our results are also qualitatively consistent with those recently obtained by Baldi et al. (2002) and 
Della Ceca et al. (2002) based on a hardness ratio analysis of {\it XMM-Newton} data where a lack
 of flat sources above a few 10$^{-14}$ \cgs~ is observed.

Overall, these results match well with the 
fact that the sources which contribute the most to the hard CXB, 
lie around the break of the Log$N-$Log$S$ distribution, i.e.  at $F_{2-10}\sim$10$^{-14}$ \cgs,
below which the slope of the Log$N-$Log$S$ slightly flattens (Cowie et al. 2002, Hasinger et al. 2001, Baldi et al. 2002a).
Finally we find no significant correlation between $\Gamma$ and $z$ for all our optically identified sources. 
\begin{figure}[h!]
\psfig{figure=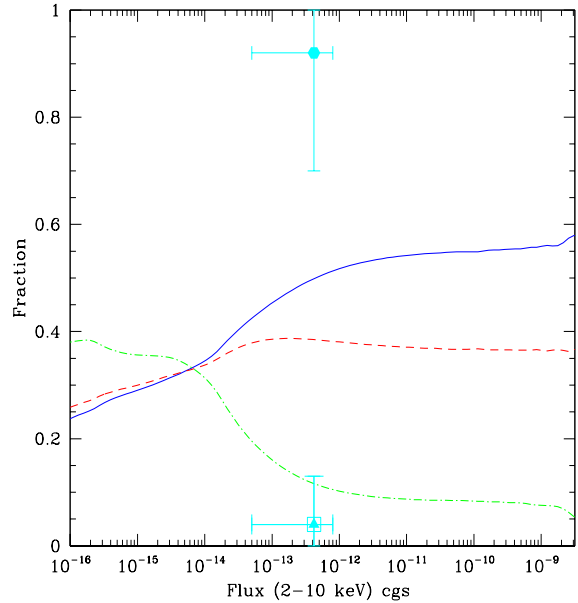,height=8.5cm,width=8cm,angle=0}
\caption{Fractions of absorbed sources in our complete subsample ({\it points}) compared to theoretical predictions ({\it lines}) of 
Comastri et al. (2001). We use the parameters resulting from the best fit model (Table 6) and, for the optically 
unidentified sources, $z$=0. The solid hexagon  and the solid line are for log$N_{H}$$<$22 cm$^{-2}$; 
the open square and the dashed line are for 22$<$log$N_{H}$$<$23 cm$^{-2}$; the 
solid triangle and the dash-dotted line for log$N_{H}$$>$23 cm$^{-2}$. 
Note that the points relative to 22$<$log$N_{H}$$<$23 cm$^{-2}$ (square) and to log$N_{H}$$>$23 cm$^{-2}$ (triangle) are coincident.
The plotted $y$-error bars correspond to 1$\sigma$ confidence level according to Gehrels (1986).}
\label{fig:xrb_best}
\end{figure}
\begin{figure}[h!]
\psfig{figure=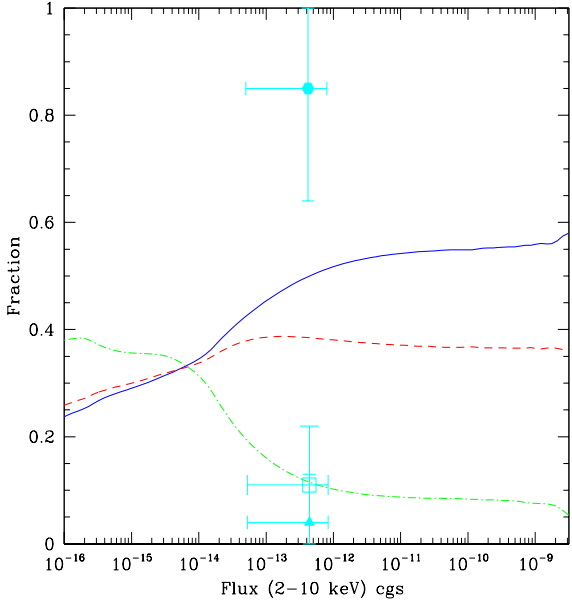,height=8.5cm,width=8cm,angle=0}
\caption{Fractions of absorbed sources in our complete subsample ({\it points}) compared to theoretical predictions ({\it lines}) of 
Comastri et al. (2001). The difference with Fig.~\ref{fig:xrb_best} is that, in this case, we force a fit with $\Gamma$=1.9 for every source and, for the optically unidentified sources in the sample, also we use $z$=1. Symbols are as in Fig.~\ref{fig:xrb_best}.}
\label{fig:xrb_19}
\end{figure}
\subsection{Absorption}
\subsubsection{Broad line X-ray absorbed objects}

Interestingly enough we find that $\sim$33\% of the sources requiring significant (at 97\% confidence level, see Table 4) 
absorption in excess of the Galactic value are broad line objects with  $N_{\rm H}$ ranging from 1 to 5 $\times$ 10$^{21}$ cm$^{-2}$.
Therefore our analysis confirms the existence of broad line AGNs which suffer from absorption in X-rays 
(Fiore et al. 2001; Georgantopoulos, Nandra \& Ptak 2001, Schartel et al. 1997, Sambruna, Eracleous \& Mushotzky 1999)
although with lower column densities compared to those found in these works. 
Also Wilkes et al. (2001) recently reported similar results for
a sample of {\it 2MASS} selected AGNs observed by {\it Chandra}: from the X-ray hardness ratio analysis they found that
the majority of type 1 objects are absorbed by intermediate column densities i.e. $N_{\rm H}\sim$10$^{21-23}$ cm$^{-2}$.
It is worth stressing that these obscured broad line objects could contribute significantly at progressively fainter fluxes 
to the fraction of the moderately absorbed sources which produces the bulk of the CXB above 2 keV (see next Section).

\subsubsection{Comparison with CXB synthesis models}
\label{comp_models}
In Fig.~\ref{fig:xrb_best} we compare the fractions of absorbed sources that we find to the
theoretical predictions of the CXB synthesis model of Comastri et al. (2001) relative to the 2-10 keV band.
For this plot we consider only sources with $F_{2-10}\geq$5 $\times$ 10$^{-14}$ \cgs~ (i.e. our completeness limit).
The column densities $N_{\rm H}$ are from the best fit model, as reported in Table 6 for each source.
The horizontal error bars associated to the points in Fig.~\ref{fig:xrb_best} correspond to our completeness limit and to 
the highest flux in the sample; the vertical error bars indicate instead the Poissonian error for each fraction value.
A mismatch between our findings and the theoretical model predictions is evident from this plot. 
The latter appear to overestimate by a factor of 2 to 10 the fraction of absorbed sources i.e. with
 $N_{\rm H}$$\geq$10$^{22}$ cm$^{-2}$,
and understimate the fraction of sources with low absorption. 
At these flux levels ($F_{2-10}$$\sim$10$^{-13}$ \cgs) the model predicts that about 55\% of the sources 
should be obscured, while we find a fraction of 20\% at most.
 
There are, however, some possible biases in our result that should be taken into account.  
The determination of the true column density for the objects without redshift is difficult and we
possibly underestimate some $N_{\rm H}$ values as $N_{\rm H}^{\rm z} \propto$(1+$z$)$^{2.6}$ (Barger et al. 2001).
To account for this effect as reported in Section~\ref{apl_s} we fitted all spectra with an absorbed power 
law model imposing $\Gamma$=1.9 (and $z$=1 for
the unidentified sources in the sample) to infer upper limits on the absorption distribution.
The results corresponding to these conservative assumptions are reported in Fig.~\ref{fig:xrb_19}. 
Also in this case we find that the fraction of absorbed sources appears to be  lower than the theoretical predicted value,
with an upper limit of about 30\%.

Another bias we consider important is 
related to the high energy roll-over of the {\it PN} effective area above 4 keV.
The effective area  decreases  by  about a factor of two between 4 and 9 keV 
while  in order to compute the theoretical predictions showed 
in Fig.~\ref{fig:xrb_best} and Fig.~\ref{fig:xrb_19} Comastri et al. (2001) assumed an uniform flat detector response.
This fact may have introduced in our analysis
a selection bias against the hardest X-ray sources.
This problem mostly affects heavily obscured sources with $N_{\rm H}\geq$10$^{22-23}$ cm$^{-2}$ which have the bulk of 
their X-ray emission above 3-4 keV i.e. exactly those sources we do not find in large quantities as foreseen by theoretical
models. 
Because of this bias it is possible that we have lost here some of the faintest, absorbed sources.
Spectral analysis of a larger number of sources could help to solve this issue.
Note, however, that even if we consider the subsample of sources 
in the harder 5-10 keV energy band
with $F_{5-10}$$\geq$5 $\times$ 10$^{-14}$ \cgs,
the mismatch between our findings and the theoretical predictions (Comastri et al. 2001) remains.
Albeit the number of sources is small (i.e. 15), we notice that our fraction of sources 
with $N_{\rm H}$$<$10$^{22}$ cm$^{-2}$ is $\sim$60-80\% compared to the Comastri et al. (2001) prediction of $\sim$35-40\%.

We also performed some simulations using {\it PN} data in order to check the possibility that 
we missed some absorbed sources (even for those belonging to the 'complete' subsample i.e. with $F_{2-10}$$\geq$5 $\times$ 
10$^{-14}$ \cgs) if their true X-ray spectra 
are the result of the combination of high column density and scattering of the primary continuum along our line of sight from
a warm medium (Turner et al. 1997). 
To test this hypothesis, we assumed as an input spectrum the average partial covering model (PCF) found in 
Seyfert 2s (i.e. $\Gamma$=1.96 and $C_{\rm f}\sim$74\%; see Turner et al. 1997, Table 19) with a typical flux of 
9 $\times$ 10$^{-14}$ \cgs~ in the 0.5-10 keV band (see Table 6). 
We found that we are able to discriminate at $>$95\% confidence level (using the $F$-test method)
between the PCF and APL models for any value of 
intrinsic $N_{\rm H}$ ranging from 10$^{23}$
down to 10$^{21}$ cm$^{-2}$.  
It is worth noting that a column density of $N_{\rm H}$=1 $\times$ 10$^{21}$ cm$^{-2}$ at $z$=0 corresponds 
to an $N_{\rm H}\leq$ 10$^{22}$ cm$^{-2}$ at $z$=1. 
Thus this effect will not change the values plotted in Fig.~\ref{fig:xrb_19}. 

Gilli, Salvati \& Hasinger (2001) presented a CXB model which assumes a luminosity
density dependent evolution (LDDE; Miyaji, Hasinger \& Schmidt 2000) model for the X-ray luminosity function (XLF) and a 
ratio $R$ absorbed/unabsorbed sources at $z$=0 ($R$=4) lower than at $z$$>$1.32 ($R$=13).
Comparing our results to their predictions (Gilli, priv. comm.),
we confirm a large mismatch for $N_{\rm H}$$<$10$^{22}$ cm$^{-2}$ and 10$^{22}$$<$$N_{\rm H}$$<$10$^{23}$
cm$^{-2}$, while for $N_{\rm H}$$>$10$^{23}$ cm$^{-2}$ we are in substantial agreement.

\begin{figure}[!]
\psfig{figure=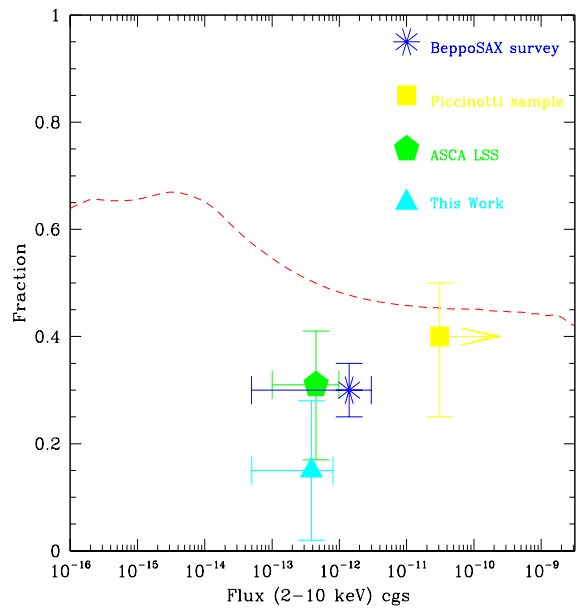,height=8.5cm,width=8cm,angle=0}
\caption{ The fraction of absorbed objects (with N$_{\rm H}>$10$^{22}$ cm$^{-2}$) as reported in the main X-ray surveys carried
 out in the 2-10 keV band with different X-ray observatories. The dashed line shows the corresponding theoretical fraction of absorbed 
objects predicted by Comastri et al. (2001).
The plotted values are from: Giommi, Perri \& Fiore (2000) ({\it BeppoSAX} 
survey, {\it asterisk}); Akiyama et al. (2000) ({\it ASCA} LSS, {\it pentagon}); this work ({\it triangle}). The value 
for the Piccinotti sample (i.e. 29 AGNs with flux limit of 3 $\times$ 10 $^{-11}$ \cgs, Piccinotti et al. 1982; {\it square}) 
is taken from Comastri (2000). 
The plotted $y$-error bars correspond to 1$\sigma$ confidence level according to Gehrels (1986).} 
\label{fig:xrb_surveys}
\end{figure}

On the other hand  some independent recent works based on {\it Chandra} and {\it XMM-Newton} observations seem to confirm our results.
Gandhi et al. (2001) have carried out a spectroscopic study of serendipitous sources at 2-10 keV fluxes fainter than ours
in some {\it Chandra} cluster observations: they reported that most of these sources are obscured by column densities of 
$\sim$10$^{21-22}$ cm$^{-2}$, with only a few showing $N_{\rm H}$$\sim$10$^{23}$ cm$^{-2}$.
Preliminary results from the {\it XMM-Newton} Bright Serendipitous Source Survey discussed in Della Ceca et al. (2002) also 
report only a few obscured sources at fluxes similar to ours.
Finally Baldi et al. (2002b) presented the 2-10 keV hardness ratio distribution of the sources belonging to the 
{\it HELLAS2XMM} medium-deep survey.
Their sample is larger than ours and also includes all sources presented here (but note their analysis is based only on the
hardness ratios rather than spectral fitting).
Their results confirm the lack of hard, and hence potentially obscured, sources with 
hard X-ray fluxes greater than $\sim$few $\times$ 10$^{-14}$ \cgs. 
For completeness we plot in  Fig.~\ref{fig:xrb_surveys} the fraction of absorbed objects 
(i.e. with $N_{\rm H}$$>$10$^{22}$ cm$^{-2}$) derived
from the main X-ray surveys carried out in the 2-10 keV band with different X-ray observatories at different limiting fluxes
and the corresponding theoretical model predictions. 
It appears that at hard X-ray fluxes between 10$^{-12}$ and 10$^{-14}$ \cgs~ the fraction of observed 
absorbed sources is considerably less than predicted by the theory.     
This figure clearly shows the inability of CXB standard theoretical models to fit the observational
data, at least for fluxes above $F_{2-10}\sim$10$^{-14}$ \cgs. This result and the mismatch emerging in the optical follow-up of the deep X-ray surveys
 (Hasinger 2002) between the predicted and the observed redshift distribution of X-ray sources suggests that CXB theoretical models should be revised (Comastri et al., in preparation).

Assuming therefore that the above observational results are real, we address here the possibility 
of modifying one (or more) of the assumptions made in standard CXB synthesis models.
As noted by Gilli, Risaliti \& Salvati (1999), the parameter space of CXB models is large and different 
combinations of input values can reproduce the CXB properties.
The critical assumptions are usually three: the intrinsic X-ray spectral shape of the sources, 
the XLFs (in particular the XLF of type 1 versus XLF of type 2 objects) and 
the column density distribution. 
All these values are already quite uncertain in the local universe and even more importantly their
evolution is unknown along $z$. It is therefore possible that 
some of the important assumptions used in theoretical models are  too rigid.
The presence of a steep ($\Gamma$$>$2) component in the soft band may not be a ubiquitous feature
of low $z$ broad line AGNs (Reynolds 1997; Matt 2000; Reeves \& Turner 2000); in particular Blair et al. (2000) suggested a possible 
evolution along $z$ of the soft excess in the {\it ROSAT} spectra of soft X-ray selected QSOs.
Furthermore high-$z$ radio quiet QSOs may have flatter intrinsic slopes than nearby ones as
found in {\it ASCA} observations (Vignali et al. 1999, Pappa et al. 2001b).

Another possibility is that QSOs may evolve in luminosity and/or in number differently than assumed in theoretical models.
For example Comastri (2000) computed the relative fraction of unobscured and obscured sources assuming a LDDE for the XLF 
instead of a pure luminosity evolution (PLE; as used in 
models plotted in Fig.~\ref{fig:xrb_best} and ~\ref{fig:xrb_19}).
The model computed in this way predicted a significantly higher number of unobscured sources than 
expected with a PLE, in better agreement with our observational results.
Also recent {\it Chandra} and {\it XMM-Newton} deep surveys require modifications to the commonly used XLFs (Hasinger 2002).

The evolution properties of type 2 sources could be different with respect to those of type 1s (see e.g. Gilli, Salvati \& Hasinger 2001).
Although the location and the dynamics of the absorption material(s) are two key parameters  with which
to understand the evolution of type 2 objects, these are only poorly known to date (i.e. Matt 2000b, Guainazzi et al. 2002).
For example if the AGNs evolution is connected to the star formation rate in their host galaxies (Oshuga \& Umemura 1999)
and/or to the growth of the central black hole (Fabian 1999), one might expect a different 
evolution in the absorption column density of type 1 versus type 2 sources. 
In particular, a faster evolution of absorbed type 2 Seyfert-like sources would reduce the above mismatch
and would also be consistent with the larger number of flat sources found by e.g. Tozzi et al. (2001) 
and Baldi et al. (2002b) at lower fluxes than ours. 

Finally another possible explanation for the discrepancy between our results and theoretical predictions
could be the existence of a new population of sources with flat intrinsic spectra emerging at faint ($\sim$10$^{-14}$ \cgs)
hard X-ray fluxes. In the next Section we discuss possible objects of this kind. 

\subsection{X-ray loud 'normal' galaxies.}
\label{dull}
Two sources in the sample, CXOU J031238.9-765134 (n.3, the so-called P3 in Fiore et al. 2000; 
$L_{2-10}$$\sim$3 $\times$ 10$^{42}$ erg s$^{-1}$) 
and XMMU J140145.0+025330 (n.41; $L_{2-10}$$\sim$2.6 $\times$ 10$^{43}$ erg s$^{-1}$ assuming $z$=0.25), 
are optically classified as normal galaxies due to the lack of
emission lines in their optical spectra, but their hard X-ray luminosities are significantly 
higher than typically found in the local galaxy population (Fabbiano 1989).
For these reasons we consider them as members of the optically dull X-ray bright galaxies class.
From our analysis both sources have a flat X-ray photon index with $\Gamma$$\sim$1 (n.3) and $\Gamma$$\sim$0.7 (n.41) when
a simple power law model (SPL) is applied. 
We also derive upper limits on the absorption column density $<$10$^{23}$
cm$^{-2}$ imposing $\Gamma$=1.9 in both sources; furthermore, we do not find any evidence of a strong iron line in their spectra, which is a typical features of reflection-dominated objects.
Comastri et al. (2002) reported the spectral energy distribution of P3 from radio to X-rays. 
On the basis of the faintness of its radio emission these authors excluded the presence of an
 advection dominated accretion flow (Narayan et al. 1998).
The SED of P3 could be consistent with a  BL Lac-type classification.
However the lack of X-ray variability and the presence of a large Calcium break 
is at odds with such a hypothesis.
Based on these overall properties Comastri et al. (2002) 
suggested that a Compton thick AGN was a very plausible explanation 
for the X-ray properties of the source. 
However, the real nature of P3 is still uncertain.   
Much less is known about source n.41, so that a detailed investigation on the nature of this object is not possible 
at the moment.

The population of these 'normal' galaxies seems to emerge at faint X-ray fluxes, well below 
10$^{-13}$ \cgs. Such sources  have not been found in previous {\it ASCA} and {\it BeppoSAX} surveys 
but this may be due to the large identification error boxes of these observatories.
Recently Barger et al. (2001) presented Keck spectroscopy of a complete sample of 20 hard X-ray 
sources with $F_{2-10}\geq$3.8 $\times$ 10$^{-15}$ \cgs~ 
detected in the {\it Chandra} Hawaii Deep Survey Field SSA13. Four galaxies in the sample ($\sim$20\%) show 
no clear sign of nuclear activity but have 2-10 keV luminosities greater than 10$^{42}$ erg 
s$^{-1}$, i.e. they are X-ray loud. As stated by these authors, these sources may be rather common X-ray emitters.
The peculiar characteristics of these sources could be explained by a very high column density 
obscuring both the broad and narrow line regions of the active nucleus or by beaming.
However, as pointed out by Hornschemeier et al. (2001), more accurate optical data are needed before definitively
ruling out the possible existence of weak emission lines.

In relation to the CXB population synthesis models, these objects are very interesting because
they 'appear' at fluxes around 10$^{-14}$ \cgs~ i.e. near the turnover of the Log$N-$Log$S$ and have a hard photon index.
The completion of multiwavelength follow-up observations of pencil beam (Rosati et al. 2002; 
Hasinger et al. 2001; Willott et al. 2001) and shallower (Baldi et al. 2001; Della Ceca et al. 2002; F02) X-ray surveys will allow 
an accurate estimate of how common these 'normal' galaxies  are and what is their relative contribution 
to the CXB.  
\section{Conclusions}
We have reported spectral results for a sample of 41 hard X-ray sources detected
serendipitously in seven {\it EPIC} fields and selected in the 2-10 keV band.
A detailed spectral analysis has been performed in order to measure source-by-source the 0.3-10 keV continuum shape,
the amount of cold (and, possibly, ionized) absorbing matter and the strength of other spectral features.
Complementary to deep pencil beam surveys, our shallower survey allows us to investigate in some detail the
spectral properties of faint serendipitous sources. This is a field of study almost unexplored with previous X-ray satellites.
We have found an average photon index  $\langle\Gamma\rangle$=1.67$\pm$0.04 using a simple power law fit with 
Galactic absorption for the whole sample. 
Considering only sources with $F_{2-10}\geq$5 $\times$ 10$^{-14}$ \cgs~ (i.e. our completeness limit) we
 obtain a $\langle\Gamma\rangle$=1.69$\pm$0.04, which is consistent with average values reported 
in recent {\it Chandra} and {\it ASCA} works at similar fluxes.

We have also shown how surveys of the kind described here can constrain some of 
the assumptions used in CXB population synthesis models 
(either spectral shape, XLF and/or column density distribution).
In particular, we found a mismatch between our observational results and those predicted by the CXB theoretical models
relative to the fractions of absorbed versus unabsorbed sources above 5 $\times$ 10$^{-14}$ \cgs~ in the range 2-10 keV.
Extremely deep pencil beam exposures do not stress this trend, very likely because of their bias towards fainter 
fluxes in the source selection.
We have also been able to collect information about unusual objects such as broad line X-ray 
obscured AGNs and optically dull X-ray bright galaxies.
We are currently analysing further {\it XMM-Newton} observations with a goal of obtaining at least 100 source 
spectra  which will allow us to put our results on more sound statistical grounds.

\begin{acknowledgements}
We thank the {\it Hellas2XMM} Team for the optical identifications. We 
thank Andrea Comastri for providing us some results from his CXB synthesis model in electronic form.
We also thank the anonymous referee for his/her suggestions that helped to improve the manuscript considerably.
We are grateful to Roberto Gilli, Alessandro Baldi, 
and Roberto Della Ceca for interesting comments. E.P. thanks Andrea De Luca and Matteo Guainazzi for
helpful discussions on data reduction procedure.  
This research has also made use of the Simbad database, operated by the Centre de Donnees Astrophisique de Strasbourg (CDS).
This work is partially supported by the Italian Space Agency (ASI). E.P. acknowledges financial support from MIUR for
the Program of Promotion for Young Scientists P.G.R.99.  
\end{acknowledgements}

 
\end{document}